%% file: main.tex
\documentclass[10pt,twocolumn,letterpaper]{article}

\usepackage[pagenumbers]{cvpr} 
\usepackage{url}

\input{preamble}
\definecolor{cvprblue}{rgb}{0.21,0.49,0.74}
\usepackage[pagebackref,breaklinks,colorlinks,allcolors=cvprblue]{hyperref}
\def\paperID{NTIRE-39} 
\def\confName{CVPR}
\def\confYear{2026}

\newcommand{\norm}[1]{\left\lVert#1\right\rVert}

\newcommand\blfootnote[1]{%
  \begingroup
  \renewcommand\thefootnote{}\footnote{#1}%
  \addtocounter{footnote}{-1}%
  \endgroup
}

\begin{document}

\title{DRIFT: Deep Restoration, ISP Fusion, and Tone-mapping}

\author{Soumendu Majee $^*$ \and Joshua Peter Ebenezer$^*$  \and Abhinau K. Venkataramanan \and Weidi Liu \and Thilo Balke \and Zeeshan Nadir \and Sreenithy Chandran \and Seok-Jun Lee \and Hamid Rahim Sheikh \\ Samsung Research America\ }

\twocolumn[{
\maketitle

\begin{center}
  \includegraphics[width=0.98\linewidth, trim={0 0.7cm 0 0.8cm}]{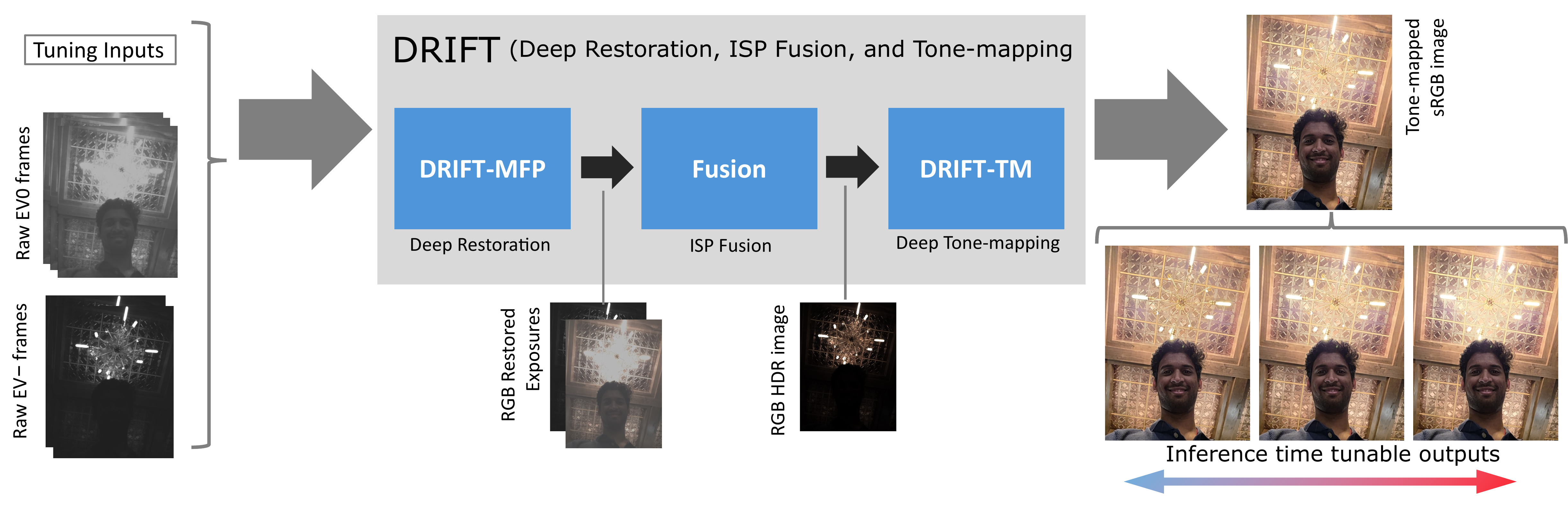}
  \captionof{figure}{Overview of the proposed Drift Pipeline. In the first part of DRIFT, DRIFT-MFP performs deep restoration of the the multi-frame raw data comprising of regular (EV0) and short (EV-) exposures and outputs single RGB restored frames for each of the two exposures. The Fusion ISP then fuses the exposures together to form a single frame HDR RGB image. Finally, DRIFT-TM performs efficient tone-mapping on the HDR RGB image to produce the final sRGB tone-mapped image. Our method allows for passing tuning inputs during inference to adjust the appearance of the final output.}
  \label{fig:proposed_pipeline}
\end{center}
}]

\input{0_abstract}    
\blfootnote{* Equal Contribution}

\input{1_intro}

\input{2_content}
{
    \small
    \bibliographystyle{ieeenat_fullname}
    \bibliography{main}
}

\end{document}

%% file: 0_abstract.tex
\begin{abstract}
Smartphone cameras have gained immense popularity with the adoption of high-resolution and high-dynamic range imaging.  As a result, high-performance camera Image Signal Processors (ISPs) are crucial in generating high-quality images for the end user while keeping computational costs low. In this paper, we propose DRIFT (Deep Restoration, ISP Fusion, and Tone-mapping): an efficient AI mobile camera pipeline that generates high quality RGB images from hand-held raw captures. The first stage of DRIFT is a Multi-Frame Processing (MFP) network that is trained using a adversarial perceptual loss to perform multi-frame alignment, denoising, demosaicing, and super-resolution. Then, the output of DRIFT-MFP is processed by a novel deep-learning based tone-mapping (DRIFT-TM) solution that allows for tone tunability, ensures tone-consistency with a reference pipeline, and can be run efficiently for high-resolution images on a mobile device. We show qualitative and quantitative comparisons against state-of-the-art MFP and tone-mapping methods to demonstrate the effectiveness of our approach.
 
\end{abstract}

%% file: 1_intro.tex
\section{Introduction}
\label{sec:intro}

Improvements in mobile camera technology continue to drive smartphone sales and have made convenient and high-quality digital photography accessible to millions of people\cite{delbracio2021mobile}. The small size of mobile cameras, while offering portability and convenience, also poses significant engineering challenges to producing high-quality photographs\cite{delbracio2021mobile,wronski2019handheld}. Typically, mobile cameras are used in a handheld manner, where the sensor captures noisy, raw frames in Bayer, Tetra (2x2 neighbors share the same color), or Hexadeca (4x4 neighbors share the same color) format. The raw frames are processed by the camera's Image Signal Processing (ISP) pipeline that fuse the raw frames, and apply a sequence of restoration and enhancements before delivering the final image to the user's gallery.

Broadly, a multi-frame camera ISP may be functionally divided into two stages. First, a Multi-Frame Processing (MFP) system takes multiple noisy raw frames as input and performs alignment, denoising, demosaicing, and optionally, super-resolution, yielding a linear RGB image as its output\cite{wronski2019handheld,tico2008multi}. Although handshake motion may introduce motion blur to input frames, multiple frames from a handheld capture contain additional information due to the availability of different views of the same scene. As a result, aligning and combining them offers benefits such as improved detail recovery, super-resolution, and high dynamic range.Camera imaging, especially high dynamic range imaging, is able to capture a wide dynamic range from real scenes. However, modern display devices and print media are limited in the dynamic range they can display. This necessitates approaches that can map a high-dynamic-range image into a low-dynamic image with desirable brightness, contrast, and color suitable for display  or print media. These demands are met by the second stage of the ISP, which is tone-mapping. Tone-mapping converts linear luminance data from the output of MFP into visually appealing images suitable for displays with limited dynamic range\cite{ma2015high}. By compressing the dynamic range, adjusting contrast and clarity, modifying colors, and reducing haze, tone-mapping aligns images with human visual perception.

In this paper, we propose DRIFT (Deep Restoration, ISP Fusion, and Tone-mapping), a unified AI-driven camera ISP solution that integrates multi-frame denoising, super-resolution, and tone-mapping into a cohesive framework. 
DRIFT-MFP and DRIFT-TM are intentionally presented together as a unified system because restoration artifacts often dominate final perceptual quality after tone-mapping, and tone-mapping design must account for restoration outputs, HDR compression, and tiling behavior. Optimizing these components independently leads to suboptimal results in practice, as illustrated in \cref{fig:codesign}.

\begin{enumerate}
    \item \textbf{Multi-Frame Denoising and Super-Resolution:} We propose a robust denoising and super-resolution network (DRIFT-MFP) that learns to process multiple frames to effectively suppress noise and improve detail. In order to achieve this, we use an efficient architecture with a stabilized GAN training objective using discriminator feature matching.
    \item \textbf{Tone-mapping Integration:} Our method seamlessly integrates tone-mapping with denoising to enhance the dynamic range and visual appeal of the final image. We propose an AI based tone-mapping architecture (DRIFT-TM) that ensures \textbf{tone-consistency} with a reference pipeline as well as \textbf{tunability} of the tone-mapping operator without retraining the AI model.
    \item \textbf{Efficiency and Scalability:} The proposed approach is designed to be computationally efficient, making it suitable for fast processing on mobile devices.
\end{enumerate}

%% file: 2_content.tex
\section{Related Work}

\subsection{Multi-Frame processing}

Early work on handheld multi-frame image alignment and denoising~\cite{wronski2019handheld,tico2008multi} relied on classical, hand-designed methods to align multiple frames, remove outliers, and blend them to achieve a denoised result. Deep-learning based methods can instead rely on priors learnt during training to denoise the image more effectively. BIP-Net~\cite{dudhane2022burst} 
used deformable convolutions in its architecture to align multiple frames to a base frame. Burstormer~\cite{dudhane2023burstormer} introduced a ``reference-based feature enrichment'' block. General purpose architectures such as Restormer~\cite{Zamir_2022_CVPR} and NAFNet~\cite{avidan_simple_2022} have been proposed to solve a wide variety of tasks, including image denoising and restoration. We chose to use NAFNet as the core architecture for our work due to its efficiency as well as its strong performance on benchmarks. NAFNets have seen wide acceptance, as evidenced by their use by winning teams in the NTIRE 2025 challenges on raw image restoration, super-resolution, and burst HDR~\cite{conde2025ntire,lee2025ntire}. Surprisingly, we found that even without using the loss functions proposed in this work, NAFNet outperforms custom architectures such as Burstormer and BIPNet on metrics. However, we do find that for images with significant global motion, Restormer performs in-network registration much better than NAFNET. This is expected as transformer architectures have global receptive fields and can hence correct global motions better. 
\subsection{Loss Functions}
Image restoration and super-resolution tasks typically use a fidelity loss ($\mathcal{L}_{data}$) such as the L1 loss or the L2 loss that encourages the network to produce images that are close to the ground-truth in the pixel space. Prior work has demonstrated the L1 loss produces sharper results than the L2 loss~\cite{goodfellow2016deep}, hence we use it in this work.

When applied to image restoration tasks, conditional adversarial training~\cite{goodfellow2014generative,isola2017image} consists of two networks - the generator (\(G\)) and the discriminator (\(D\)) - playing the following minmax game:
\begin{align}
    \min\limits_{G} \max\limits_D &E_{x,y}\left[\mathcal{L}\left(G\left(x\right), y\right) + \log D\left(G\left(x\right)\right)\right] + \nonumber \\ &E_{x,y}\left[\log \left(1 - D\left(G\left(x\right)\right)\right)\right],
\end{align}
where \(x\) and \(y\) denote the input and ground-truth respectively, and \(\mathcal{L}\) denotes a ``reconstruction loss'' between the generator's output and the ground truth.

Perceptual losses~\cite{johnson2016perceptual} are based on an empirical observation that internal feature layers of a deep network (VGG16 or VGG19~\cite{simonyan2014very}) trained to perform image classification on ImageNet (a large image database)~\cite{imagenet} encode high-level perceptual and semantic information about images. 
Building on the foundations of the VGG perceptual loss, LPIPS~\cite{zhang2018unreasonable} trained a linear weighting of deep feature activation differences to predict the perceived differences between images, as determined by a user study. However, use of this perceptual loss often results in artifacts~\cite{fu2022edge, Krawczyk_2023}. 

We posit that artifacts generated by perceptual losses may be attributed, in part, to the domain gap between image classification (which VGG was trained for) and the raw image restoration task~\cite{pihlgren2023systematic}. This domain gap may lead to artifacts. In addition, since the VGG network is frozen during generator training, optimization with respect to LPIPS may converge prematurely if the network's output and ground-truth induce similar activations in the pretrained VGG network. Based on these hypotheses, instead use discriminator feature matching \cite{pix2pixhd,salimans} for the first time in a raw multi-frame image restoration task.

\subsection{Tone-mapping}
Tone-mapping is an essential component of a mobile image processing pipeline that converts a high dynamic range (HDR) linear image with high bit-depth into a lower dynamic range image that can be displayed on a commercial device. \cite{cerda2018tone, delbracio2021mobile, ma2015high}.

Various classical tone-mapping operators have been proposed in the literature~\cite{mantiuk2008display, reinhard2021high, gastal2011domain, petit2013assessment, mertens2007exposure}. 
Typically, tone-mapping is performed as a sequential application of various blocks relating to dynamic range compression, contrast enhancement, etc.
In order to achieve the best tone-mapping quality, it is important to run the various tone-mapping blocks at full-resolution, high bit-depth, or more scales in multi-scale decomposition based blocks.
This creates a challenge in running high-fidelity tone-mapping pipelines on mobile devices while balancing run-time and memory requirements. 
With advances in deep learning, there has been a growing interest in the development of tone-mapping 
operations using deep learning \cite{yang2024learning, guo2021deep, zhang2023lookup} based on semi-supervised \cite{wang2022learning, guo2021deep} and supervised learning \cite{kinoshita2019convolutional, tmogan, llflut, rana2019deep}. 
However, a fundamental challenge with existing deep learning (DL) based methods is the computational cost and lack of 
options to tune different aspects of the tone such as local contrast, brightness, and HDR region recovery \cite{tmogan, llflut}. 
Additionally, when input images are presented to these tone-mapping operators in tiled form (e.g., to save memory in high resolution 
imaging pipelines), they do not guarantee consistent tone across different input tiles. 

 In this work, we solve these challenges by proposing a novel deep network to perform tonemapping in an efficient manner. The solution also allows for tunability after deployment, and we incorporate global information into the network that prevents it from generating tiling artifacts.

\begin{figure}[t]
 \centering
  \includegraphics[width=0.9\linewidth]{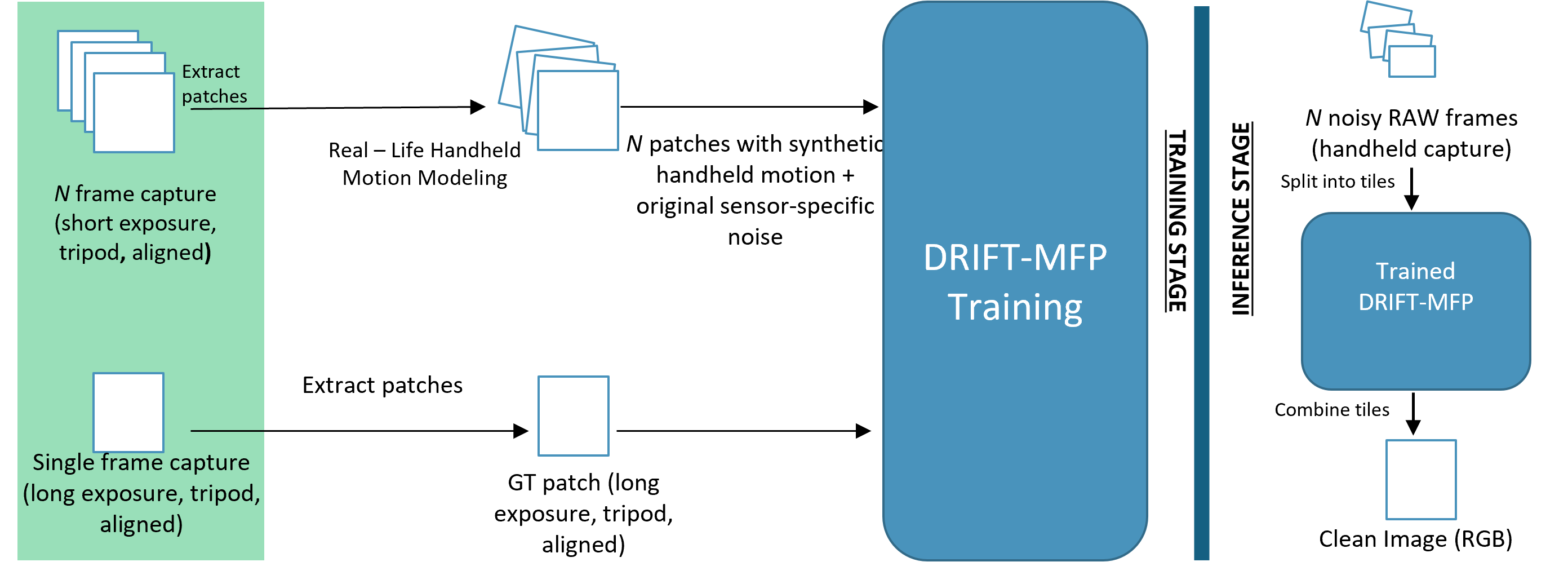}
 \caption{Overview of the training and inference pipelines for DRIFT-MFP. During training, we model handshake motions using homographies from real data and apply them to tripod captures of corresponding short exposure and long exposure images with equalized brightness.}
 \label{fig:training_MFP}
 \end{figure}

 \begin{figure}[t]
 \centering
  \includegraphics[width=0.5\textwidth]{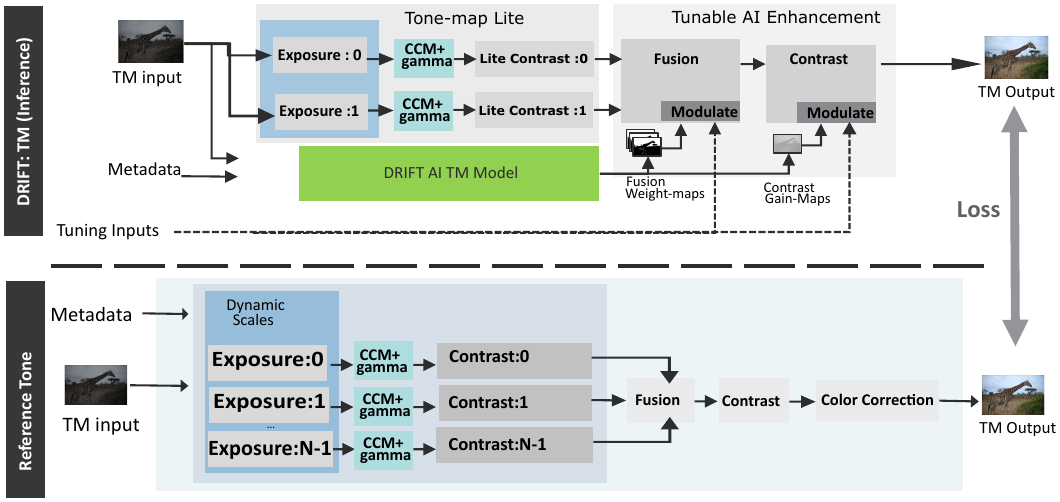}
 \caption{Overview of the training and inference pipelines for DRIFT-TM. We train DRIFT-TM using a computationally expensive referenece tone-map. DRIFT-TM learns the residual enhancements from a light-weight tonemap allowing more robust learning as well as tunability by modulating the enhancements.}
 \label{fig:aiutm_total_arch}
 \end{figure}

 \begin{figure}[t!]
 \centering
  \includegraphics[width=0.50\textwidth]{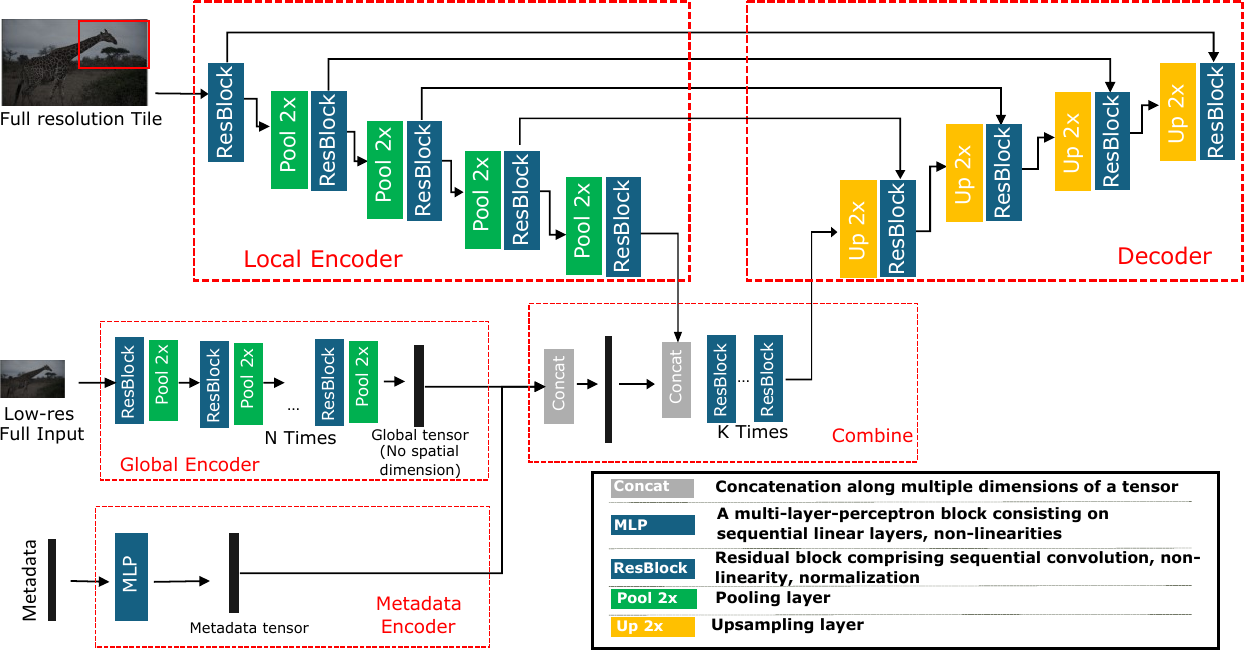}
 \caption{DRIFT Tone-map network architecture. We incorporate a local encoder that encodes a full-resolution image per tile, a global encoder that encodes a low-resolution full image, and a metadata encoder to encode capture metadata.}
 \label{fig:aiutm_ai_arch}
 \end{figure}

\section{Multi-Frame Denoising and Super-Resolution}
\label{sec:MFP}

\subsection{Network Architecture}
Inspired by their use in NTIRE 2025 challenges \cite{conde2025ntire,ntire2025night}, we use a NAFNet~\cite{avidan_simple_2022} as the core multi-frame processing architecture. The NAFNet was configured to accept 11 RGB noisy frames (33 total channels) as inputs and output a single RBG image. NAFNets are free of non-linear activation functions and use simple normalization and convolution layers, which make them highly suited for efficient on-mobile implementation. For example, the deformable convolutions used by networks such as Burstormer are not natively supported currently on Snapdragon NPUs \cite{snpe}, and transformer blocks in networks such as Restormer are prohibitively slow. To support super-resolution, all input low-resolution RGB frames were upscaled using Bilinear interpolation before passing to the NAFNet. The same approach was used with all other baseline networks discussed in Sec. \ref{sec:results_MFP}. In this manner, the networks were adapted to super-resolution without modifying their core architectures. For adversarial training, we used a Patch-GAN discriminator \cite{isola2017image}.
\subsection{Dataset}
We collected a dataset of 1300 paired input and ground truth 12MP images on a tripod using a Samsung Galaxy S24 Ultra, following~\cite{madhusudana2024mobile, khan2025mfsr,ANAYA2018144,ponomarenko_image_2015,xu_multi-exposure_2022}. The inputs were captured as 11 frames with exposure times and ISOs determined by the camera's autoexposure. The ground truth was created by blending (averaging) and demosaicing 11 ``long-exposure'' frames. The ISO of the long exposure images was fixed at 50 and their exposure was automatically calculated such that they matched the brightness of the input images. We refer to them as ``long exposure'' since the low ISO value typically necessitates a longer exposure time compared to that of the inputs.

We then collected a dataset of 1200 11-frame handheld captures and computed the global homographies between each frame and the first (``reference'') frame following~\cite{khan2025mfsr}. These homography matrices model real human hand-shake motion, which is absent in tripod captures. By applying real homographies on tripod captured data, we synthesize realistic hand-motion. Notably, while~\citet{khan2025mfsr} sampled homography matrices independently at random for each input frame, we modeled temporal correlations in handshake motion by sampling groups of 10 homographies from the handheld capture set and applying them to the 10 non-reference frames in each input burst. In this way we create a dataset of noisy Bayer frames as inputs, and a clean RGB image (aligned with the input reference frame) as ground truth, with accurate sensor noise characteristics and simulated handheld motion. In addition, for the super-resolution task, input frames were bilinearly downscaled by a factor of 4\(\times\) as in prior work~\cite{Bhat_2021_CVPR}.


\begin{figure*}[ht!]
 \centering
  \includegraphics[width=\textwidth]{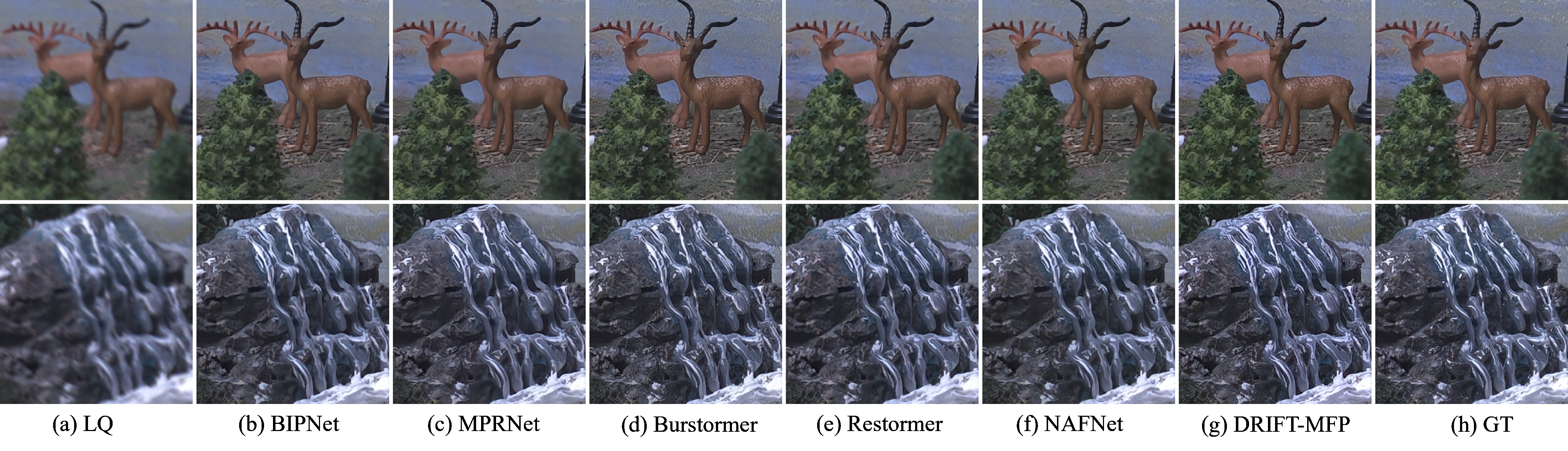}
 \caption{ Denoising results across various scenes. For each scene (row), the columns correspond to: (a) The low-quality input image, (b) BIPNet, (c) MPRNet (d)Burstormer, (e) Restormer (f) NAFNet (g) Our proposed method DRIFT-MFP, and (h) The Ground Truth (GT). Our method consistently outperforms the baselines in terms of visual quality and fidelity to the GT. }
 \label{fig:mfp_results}
 \end{figure*}
\begin{figure*}[ht!]
 \centering
  \includegraphics[width=\textwidth]{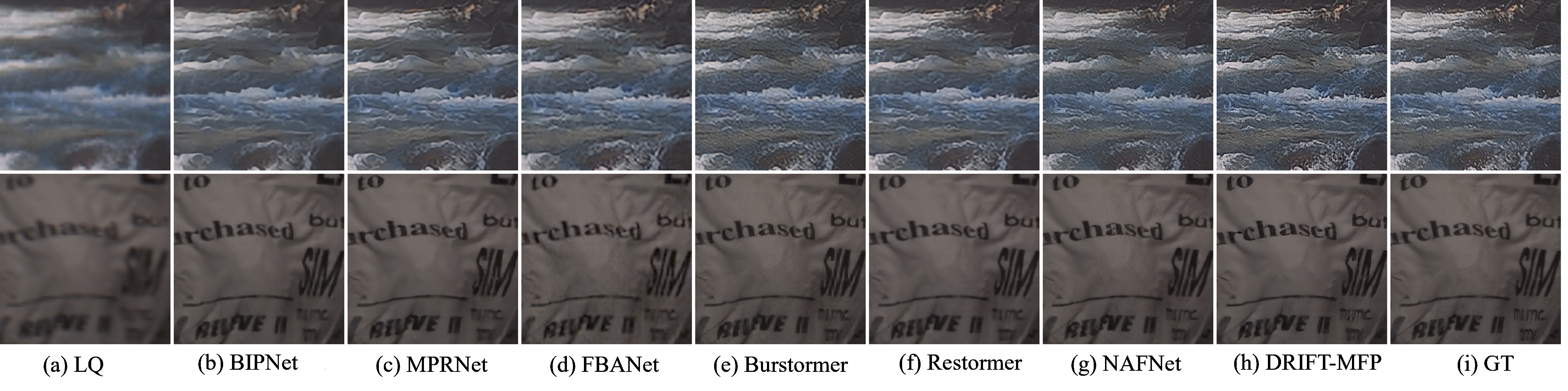}
 \caption{4x SR results across various scenes. NAFNET trained with LPIPS achieves the best FID score but produces artifacts as seen in the second row while our method maintains fidelity to GT without generating unnatural artifacts.}
 \label{fig:SR_results}
\end{figure*}

\subsection{Adversarial Perceptual Loss}
\label{sec:APL}
Unlike a pretrained frozen VGG network, a discriminator network continually adapts during training (due to the backpropagated discriminator loss) to be more sensitive to differences between the generator's outputs and ground truth images. Hence, unlike a pretrained classification network, internal feature activations of the discriminator are expected to be relevant to the restoration task at hand, and can be introduced as a training objective to stabilize GAN training and increase realism~\cite{salimans,pix2pixhd}. Since post-activation features  $\sigma(F_{D_i})$ may be zeroed out (e.g. ReLU activations) or suppressed (e.g. Sigmoid activations) to compress information for subsequent network stages, and has been shown to be less informative for image restoration tasks\cite{wang2018esrgan}, we use the difference between pre-activation features at each layer to compute the loss metric, which we term the ``Adversarial Perceptual Loss'' (APL).

\begin{equation}
  \mathcal{L}_{APL} = \sum_{i=1}^N \norm{F_{D_i}\left(G\left(x\right)\right) - F_{D_i}\left(y\right)}_1, 
\end{equation}
where $F_{D_i}$ denotes the feature representation of the discriminator $D$ at stage $i$.

Hence, the overall optimization objective for the generator is  
\begin{equation}
  \mathcal{L}_G = \mathcal{L}_{data} + \lambda_1 \mathcal{L}_{GAN} + \lambda_2 \mathcal{L}_{APL}.
\end{equation}

\subsection{Fusion}
\label{sec:Fusion}
We capture an EV- frame with an exposure of 1/8th that of the EV0 frames to enable HDR imaging. It is aligned to the EV0 reference frame using homography estimation \cite{fischler1981random}. 
We follow Mertens et al.\cite{mertens2009exposure} and \cite{burt1983multiresolution} to fuse the restored long-exposure RGB image with the short exposure image.
Any local-motion based mismatches between the exposures are handled by a de-ghosting step.

\section{Tone-mapping}
\label{sec:UTM}

We introduce a novel deep learning-based tone-mapping architecture that ensures tone-consistency with a reference pipeline, tunability, and efficient high-resolution processing. Unlike conventional DL tone-mapping methods that predict the final RGB image directly, our architecture predicts residual enhancements to a lightweight baseline tone-map. This simplifies the network's task and leverages functional decomposition: the baseline handles global color and exposure, while the network refines contrast and HDR details. Additionally, our method allows tunable enhancement at inference time, making it suitable for mobile deployment. The overall training and inference pipelines are illustrated in \cref{fig:aiutm_total_arch}.

\subsection{Reference Tone-map}
To achieve the desired characteristics in brightness, contrast, and color reproduction, we first built a fine-tuned computationally complex non-DL reference tone-map pipeline. This is shown in the bottom half of \cref{fig:aiutm_total_arch}. The reference tone-mapping pipeline utilizes synthetic exposure fusion~\cite{mertens2007exposure, yang2025high, li2019hybrid}  combined with various color, contrast, and brightness adjustment blocks to create the individual exposures. In addition, the method consists of blocks to adjust contrast and brightness after exposure fusion. 

\subsection{Tone-map Lite}
To enable efficient inference-time tonemapping, we use ``Tone-map Lite'', a non-DL algorithm that computes a light-weight tone-map with a desired reference processing order. Tone-map Lite is designed to be a light-weight version of the reference tone-map that can match the subjective ``feel'' of the reference tone-map in terms of approximate brightness and color. Specifically, Tone-map Lite predicts two tone-mapped images that approximately correspond to the darkest and brightest frames generated in the reference tone-map for exposure fusion.

The two outputs of Tone-map Lite, which correspond to dark and bright synthesized exposures, may be denoted by $\left(S_0^Y, S_0^C\right)$ and $\left(S_1^Y, S_1^C\right)$ respectively, where the superscripts $\left(Y, C\right)$ denote the image in the luma and chroma channels respectively.

\subsection{Network Architecture}
\cref{fig:aiutm_ai_arch} describes the DRIFT-TM network architecture. The network takes a linear HDR image as input and outputs fusion weight and gain maps as outputs, which are used to enhance the output of Tone-map Lite to match the reference tone-map.  Since tone-map computation depends both on local and global image content, our architecture incorporates separate encoders that capture global and local image features. The global encoder processes a low-resolution version of the full input image to generate global features. This operation is performed once per image. The local encoder may process the full-resolution image in tiles, since memory may be limited on mobile devices. This operation is performed once for each tile. 
The global encoding allows us to get global image information even for light-weight network architectures with limited receptive field: this
prevents tile-to-tile tone inconsistency and improves robustness.
The network also inputs capture metadata that is processed through a separate metadata encoder that is run once per a full-frame image. Categorical metadata entries such as sensor/pipeline type are encoded with one-hot encoding. Quantitative metadata values such as ISO, exposure-time are normalized before encoding. 
Since sensor type and capture lighting conditions have a big impact on the noise levels and the desired enhancement level in tone-map, this approach enables us to train one network spanning various capture conditions and sensors types.

\subsection{Tunable Tone Enhancement}
\label{sec:tunableENh}
The weight and gain maps output by the network are modulated based on a tuning inputs to fine-tune the final output as desired without retraining the network. The weight/gain map outputs of the network are the Luma fusion weights: $W^Y$, the Chroma fusion weights: $W^C_0, W^C_1$, and the Luma contrast gain-map: $G$. 

Let $G_\phi : [0,1] \to [0,1]$ denote a lookup-table (LUT) used to modulate fusion weights. By applying it on the model's output, modulated fusion weights are given by
\begin{equation}
    \tilde{W^Y} = G_\phi\ (W^Y) .
\end{equation}

Similarly, let $H_{\theta_0} : [0,1] \to [0,1]$ and $H_{\theta_1} : [0,1] \to [0,1]$ denote LUTs used to modulate the synthesized exposures generated by Tone-map Lite. Hence, the modulated exposures are given by
\begin{align}
    \tilde{S_0^{Y}} &= H_{\theta_0} (S_0^Y ), \\
    \tilde{S_1^{Y}} &= H_{\theta_1} (S_1^Y ) .
\end{align}
The output of fusion in YCbCr space is given by:
\begin{align}
    I_Y &=  \tilde{W^Y} \odot  \tilde{S_0^{Y}} + (1-\tilde{W^Y}) \odot \tilde{S_1^{Y}}, \\
    I_C &= W^C_0 \odot S_0^C + W^C_1 \odot S_1^C,
\end{align}
where $I_Y$ and $I_C$ denote the image in the luma and chroma channels respectively.
The final contrast-enhanced tone-mapped image in YCbCr space is given by:
\begin{align}
    \tilde{I}_Y &=  I_Y \odot \tilde{G} , \\
    \tilde{I}_C &= I_C 
\end{align}
where $\tilde{I}_Y$ and $\tilde{I}_C$ denote the image luma and chroma channels respectively, $\odot$ denotes a point-wise multiplication, and $\tilde{G}$ denotes the modulated contrast gain-map given by
\begin{equation}
    \tilde{G}=(1-S)+S \odot G ,
\end{equation}
where $S$ denotes a point-wise strength map. The point-wise strength-map can be increased for all pixels in order to increase local contrast everywhere in the image or $S$ can be adjusted on certain parts of the image based on semantic content. Finally, both outputs \((I_Y, I_C)\) and \((\tilde{I}_Y, \tilde{I}_C)\) are converted to the RGB domain images \(I\), \(\tilde{I}\). The ability to independently choose the LUTs \(G_\phi\), \(H_{\theta_0}\), and \(H_{\theta_1}\), and the strength map \(S\) at inference-time without retraining the model enables tunable tonemapping in DRIFT-TM. We use identity mappings as default initializations for the LUTs as well as to obtain our results in the following section.

\subsection{Loss Function}
During training, we run the reference tone-map twice: once with contrast enhancement blocks off and once with them turned on. This gives us two different ground truth targets: let us refer to them as $y^0$ and $y^1$ respectively. The loss is calculated between the two ground truths and the two outputs with and without contrast enhancement: 
\begin{equation}
    \mathcal{L} = \sum_{t \in \{L_1, SSIM \}} \mathcal{L}_t\left(I, y^0\right) + \mathcal{L}_t(I, y^1).
\end{equation}
Minimizing the loss with the two ground truths $y^0$ and $y^1$ allow the network to effectively learn the orthogonal operations of HDR-control/brightness-adjustment and local-contrast adjustment.
As a consequence, modulating the weight and gain maps using the tuning LUTs can specifically target HDR-control/brightness-adjustment and local-contrast adjustment.

\subsection{Dataset}
We gathered 2000 handheld 12MP raw captures using Samsung Galaxy S25 and S25 Ultra across various light levels. Each raw frame was processed to generate an HDR linear RGB image, which was then passed through the reference tone-map pipeline to create input-ground-truth pairs. For DRIFT-TM training, 512x512 patches were randomly extracted from the full 12MP images.
 \begin{table}[b]
 \small
 \begin{tabularx}{\columnwidth}{lXXXX}
\toprule
    \textbf{Method} & \textbf{LPIPS} & \textbf{FID} & \textbf{PSNR} & \textbf{SSIM} \\ \hline

    BIPNET~\cite{dudhane2022burst} & 0.09 & 11.68 & 36.32 & 0.97 \\ 
    Burstormer~\cite{dudhane2023burstormer} & 0.04 & 6.18 & 37.06 & 0.98 \\ 
    MPRNet~\cite{Zamir_2021_CVPR} & 0.08 & 12.40 & 36.76 & 0.97 \\ 
    Restormer~\cite{Zamir_2022_CVPR} & 0.05 & 8.49 & 36.36 & 0.97 \\ 
    NAFNet \scriptsize{(L1)}& 0.10 & 29.00 & 35.48 & 0.90 \\ 
    NAFNet \scriptsize{(L1+GAN)}  & 0.15 & 45.42 & 34.01 & 0.88 \\ 
    NAFNet \scriptsize{(L1+GAN+LPIPS)}  & 0.04 & 6.230 & 37.55 & 0.97 \\ 
    DRIFT-MFP  & 0.05 & 10.734 & 37.49 & 0.97 \\

\bottomrule
\end{tabularx}
\caption{Quantitative Evaluation of Multi-frame Denoising}\label{tab:bright_comparison}
\end{table}

\begin{table}
\small
 \begin{tabularx}{\columnwidth}{lXXXX}
\toprule
    \textbf{Method} & \textbf{LPIPS} & \textbf{FID} & \textbf{PSNR} & \textbf{SSIM} \\ \hline

    BIPNET~\cite{dudhane2022burst} & 0.16 & 21.91 & 34.79 & 0.94 \\ 
    FBANet~\cite{wei2023towards} & 0.19 & 21.60 & 32.05 & 0.94 \\ 
    Burstormer~\cite{dudhane2023burstormer} & 0.09 & 12.27 & 35.07 & 0.95 \\ 
    MPRNet~\cite{Zamir_2021_CVPR} & 0.13 & 21.53 & 35.19 & 0.95 \\ 
    Restormer~\cite{Zamir_2022_CVPR} & 0.12 & 16.28 & 35.25 & 0.95 \\ 
    NAFNet \scriptsize{(L1)} & 0.11 & 17.82 & 37.75 & 0.96 \\ 
    NAFNet \scriptsize{(L1+GAN)} & 0.11 & 17.42 & 37.83 & 0.96 \\ 
    NAFNet \scriptsize{(L1+GAN+LPIPS)} & 0.07 & 8.00 & 37.00 & 0.95 \\ 
    DRIFT-MFP & 0.10 & 20.84 & 36.22 & 0.94 \\

\bottomrule
\end{tabularx}
\caption{Evaluation of Multi-frame 4x Super-Resolution}\label{tab:srbright_comparison}
\end{table}

\begin{table}[ht]
\small
\begin{tabularx}{\columnwidth}{lcccc}
\toprule
             & \textbf{NAFNET} & \textbf{Same} & \textbf{DRIFT-MFP} \\ 
    \hline
    User Preference          & 28.2\%           & 8.8\%           & \textbf{63.0\%}           \\ 
\bottomrule
\end{tabularx}
\caption{Results of user study on DRIFT-MFP denoising and super-resolution}\label{tab:user_study}
\end{table}

\section{Results}

\subsection{Multi-Frame Denoising and Super-Resolution}
\label{sec:results_MFP}

We compare our method against prior art on multi-frame denoising and super-resolution in \cref{tab:bright_comparison} and \cref{tab:srbright_comparison} on a held-out test set of 150 12MP images. For both tasks, we retrained all baseline networks on the same training dataset as DRIFT-MFP for a fair comparison.
Qualitative comparison between our method and several baselines is shown in \cref{fig:mfp_results} and \cref{fig:SR_results}. The NAFNet results shown in the Figures correspond to the network trained using GAN and LPIPS losses. As shown, particularly in Fig. \cref{fig:SR_results} (g), we find that the LPIPS loss term creates a characteristic grid-like pattern artifact which boosts its FID and PSNR scores, but is visually unpleasant. Another example is shown in the supplementary material. In order to validate this, we conducted a user study on 60 random crops of images from the test set of the multi-frame denoising and super-resolution task, since it is a super-set of the two DRIFT-MFP tasks discussed here. Eleven image quality experts participated in the user study. A majority of responses favored DRIFT-MFP, and the general feedback on the study was that the grid-like pattern can appear as detail when masked by high-texture regions, but can be very distracting on flat areas. Results of the user study are shown in \cref{tab:user_study}.

\subsection{Tone Mapping}
\label{sec:results_TM}

\subsubsection{Non-Reference Comparison}
\label{sec:results_TM_non_ref}

\begin{figure}[ht!]
    \centering
    \includegraphics[width=0.5\textwidth]{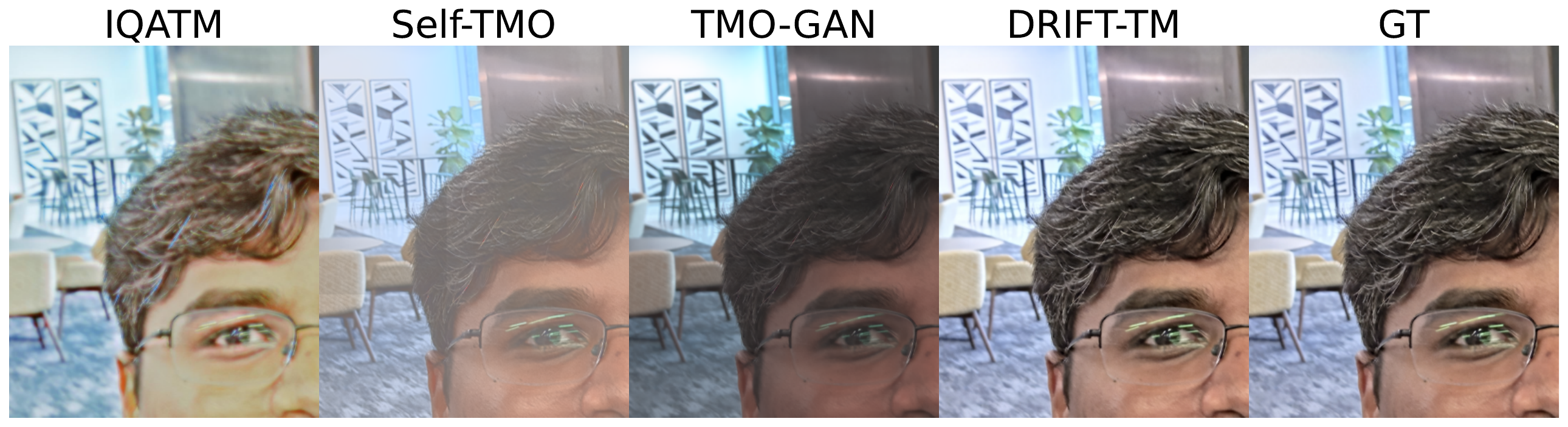}
    \label{fig:image98}
    \caption{Non-reference tone-mapping methods comparisons.
    Our method consistently outperforms the other ones in visual quality. 
    }
    \label{fig:unsup-prior_comp}
\end{figure}

In this section we compare our method against other state-of-the art methods in tone-mapping that have been pre-trained: IQATM~\cite{guo2021deep}, Self-TMO~\cite{wang2022learning}, and TMO-GAN~\cite{tmogan}.
Since Self-TMO~\cite{wang2022learning} and TMO-GAN~\cite{tmogan} were not able to run on our hardware on the full 12MP resolution images, we implemented a 4x4 tiling processing with 50 pixels overlap regions. 
In \cref{fig:unsup-prior_comp} we evaluated the methods using the TMQI~\cite{yeganeh2013Objective} Tone Mapped Image Quality Index which provides an overall quality index, TMQI-Q, a structural fidelity index, TMQI-S, and a statistical naturalness index, TMQI-N. 
\cref{fig:unsup-prior_comp} shows qualitative results of each method, while \cref{tab:reference_results} shows quantitative metrics. From this analysis, we observe that DRIFT-TM and the reference tonemap (which provides ground-truth) achieve highest subjective quality.

\begin{figure}[ht!] 
\centering 
        \includegraphics[width=1.0\linewidth, trim={0 1.2cm 0 1.2cm}]{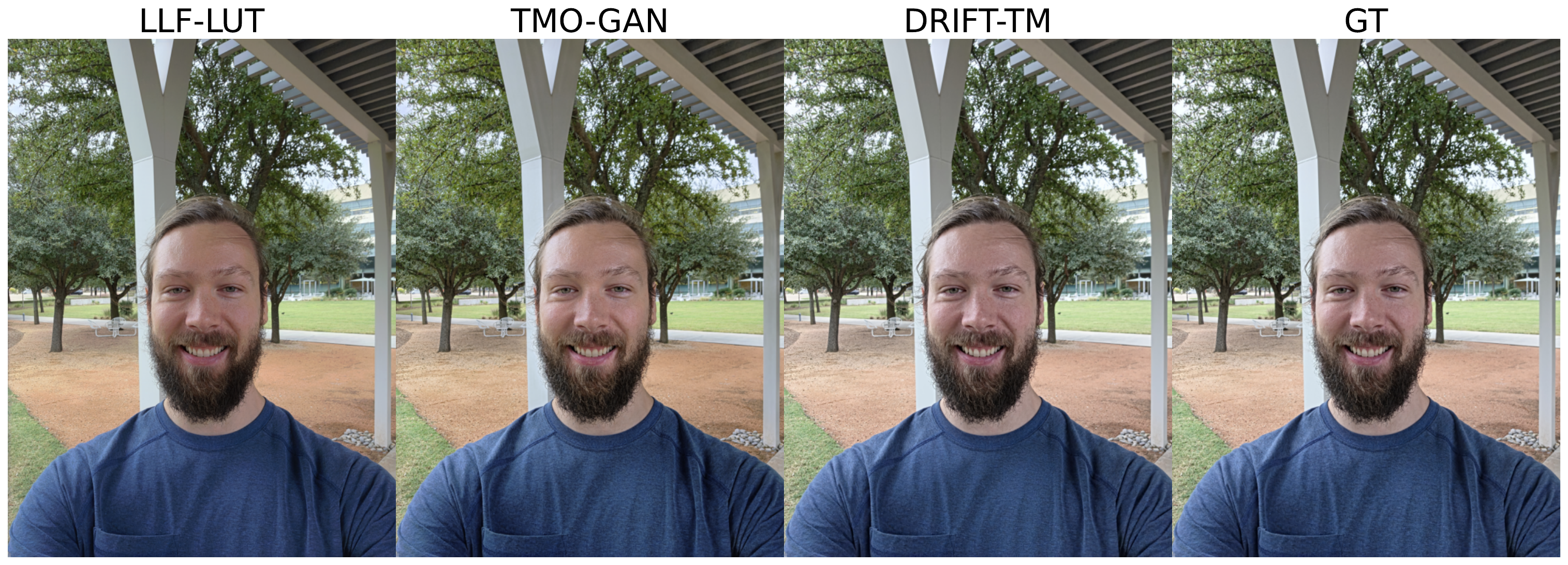}
        \label{fig:prior_comp_image3}
        \includegraphics[width=1.0\linewidth, trim={0 1.2cm 0 1.2cm}]{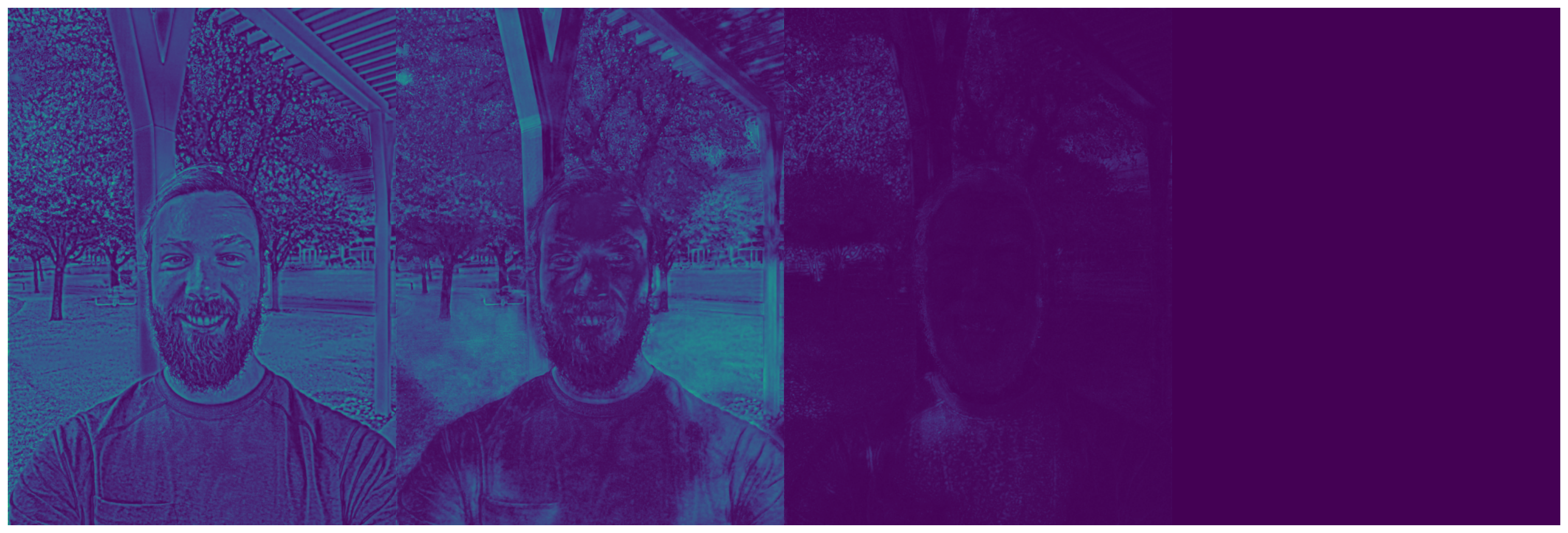}
        \label{fig:prior_comp_image4}
\caption{Visual comparison with state-of-the-art supervised learning based tone-mapping methods (top: image, bottom: difference map from GT). 
The results indicate that the both LLF-LUT and TMO-GAN have an inferior ability to match the tone-mapping operation to the GT compared to our proposed method.
}
\label{fig:prior}
\end{figure}

\begin{table}[ht]
\small
 \begin{tabularx}{\columnwidth}{lcccc}
\toprule

    \textbf{Method}                    & \textbf{TMQI-Q} & \textbf{TMQI-S} & \textbf{TMQI-N} \\ 
    \hline
    IQATM~\cite{guo2021deep}           & 0.805           & 0.713           & 0.330           \\ 
    Self-TMO~\cite{wang2022learning}   & 0.759           & 0.725           & 0.091           \\ 
    TMO-GAN~\cite{tmogan}              & 0.801           & 0.757           & 0.253           \\ 
    \textbf{DRIFT-TM}                  & \textbf{0.845}  & \textbf{0.791}  & \textbf{0.421}  \\ 
    \textbf{GT}                        & \textbf{0.847}  & \textbf{0.792}  & \textbf{0.432}  \\ 
\bottomrule
\end{tabularx}
\caption{Quantitative results of the non-reference comparison experiments with the TMQI metric.}\label{tab:nonref_comparison}
\end{table}

\subsubsection{Reference comparisons}
\label{sec:results_TM_ref}

In \cref{tab:reference_results}, we compare DRIFT-TM and state-of-the-art deep-learning (DL) tone-mapping methods in their ability to match the reference tone-map. All methods were trained on the same dataset to match the reference tone-map output and evaluated on a held-out test set of 100 images.
The evaluation metric was the similarity with the GT/reference tone-map in terms of PSNR and SSIM.
\cref{tab:results_TM_ablation} shows the quantitative metrics for each method:
DRIFT-TM shows superior values in both PSNR and SSIM metrics indicating it can learn to match the reference in a more robust manner.
Qualitative comparisons are illustrated in \cref{fig:prior}.

\begin{table}[ht]
\small
 \begin{tabularx}{\columnwidth}{lcccc}
\toprule
    \textbf{Method} & \textbf{PSNR} & \textbf{SSIM}  \\ \hline
   
    LLF-LUT (\cite{llflut}) & 30.89 & 0.95 \\
    TMO-GAN (\cite{tmogan}) & 30.83 & 0.96 \\
    
    DRIFT-TM \scriptsize{(w/o maps)}  & 34.06 & 0.98 \\
    DRIFT-TM \scriptsize{(w/o global and meta)} & 39.66 & 0.99 \\
    DRIFT-TM \scriptsize{(w/o meta)}  & 40.43 & 0.99 \\
    
    \textbf{DRIFT-TM}  & \textbf{40.59} & \textbf{0.99} \\

\bottomrule
\end{tabularx}
\caption{Quantitative results comparing the prior methods with our proposed method and its ablation study variations.}
\label{tab:reference_results}
\end{table}

\begin{figure}[ht] 
    \centering 
        \includegraphics[width=0.8\linewidth, trim={0 1cm 0 1cm}]{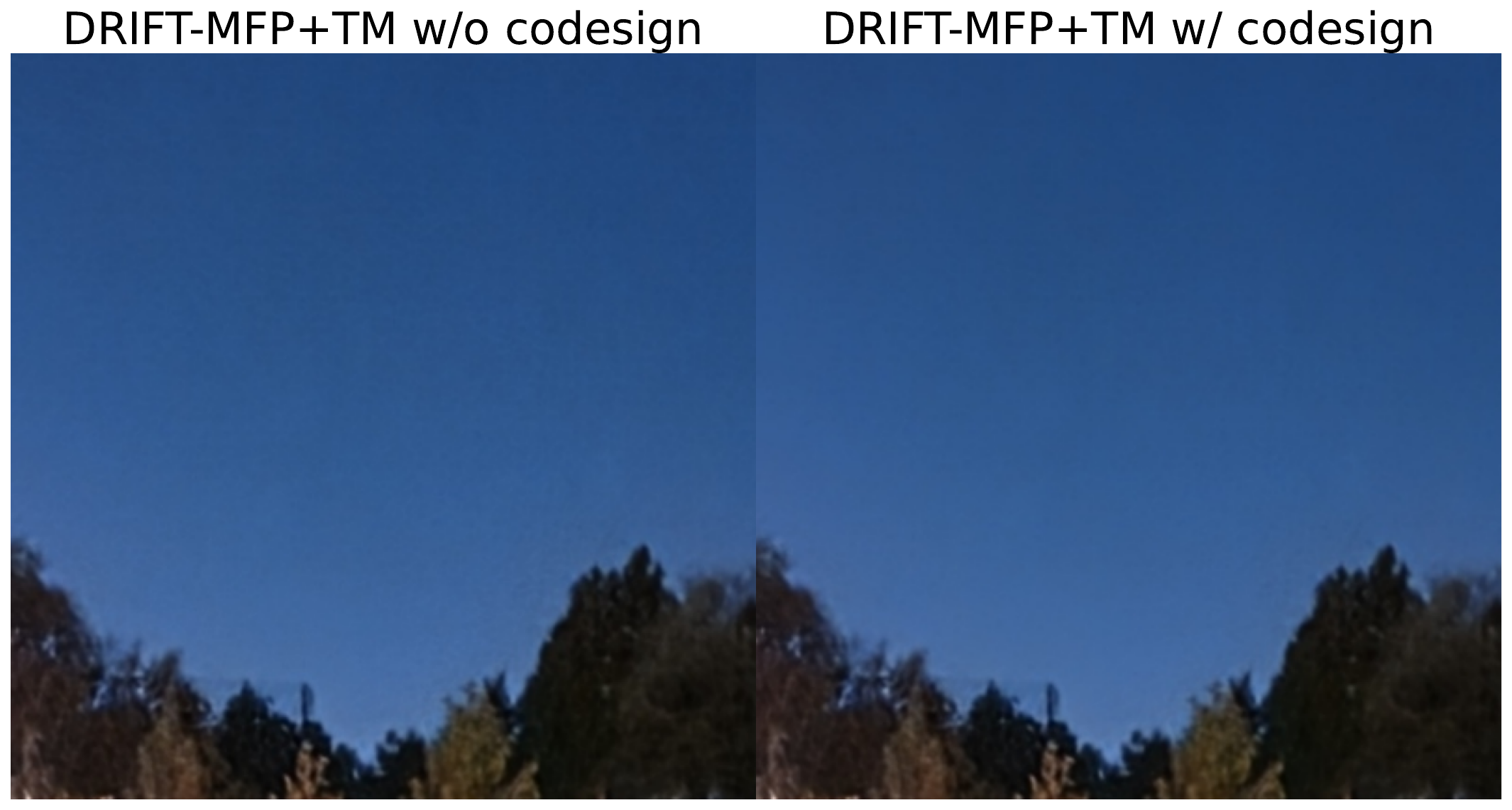}
        \label{fig:codesign_comp_image3}
\caption{Illustration showing the value of co-designing DRIFT-MFP and DRIFT-TM. 
Co-designing allows us to adjust the contrast-enhancement and rendering strengths based on the noise levels in the input and DRIFT-MFP output resulting in reduced noise and artifact enhancement in the sky regions as shown.
}
\label{fig:codesign}
\end{figure}
\subsubsection{Ablation Study}
\label{tab:results_TM_ablation}

\begin{figure}[t] 
    \centering 
        \includegraphics[width=0.9\linewidth, trim={0 1cm 0 1cm}]{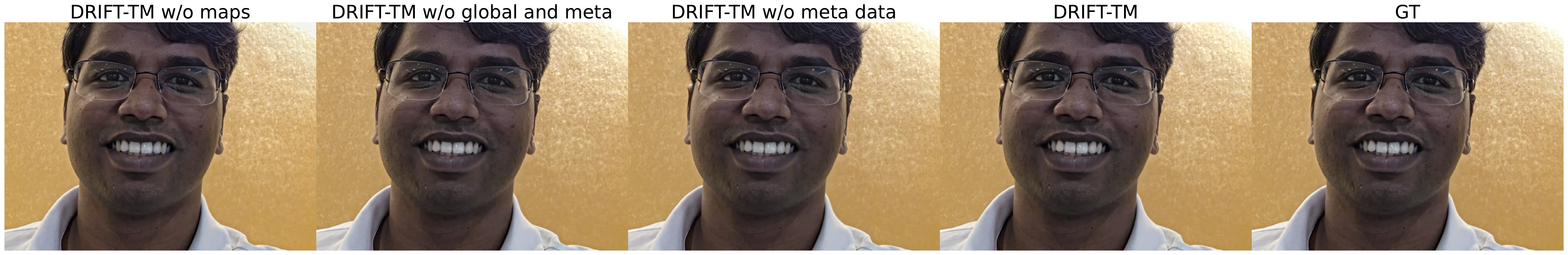}
        \label{fig:ablation_comp_image1}
        \includegraphics[width=0.9\linewidth, trim={0 1cm 0 1cm}]{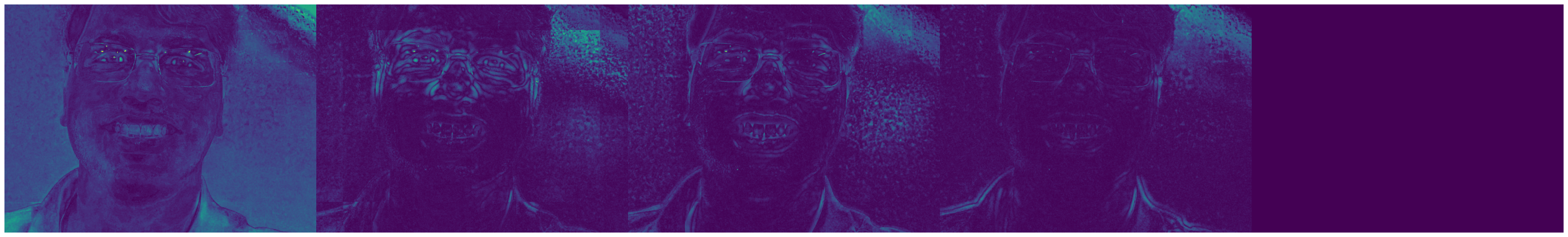}
        \label{fig:ablation_comp_image2}
        \includegraphics[width=0.9\linewidth, trim={0 1cm 0 1cm}]{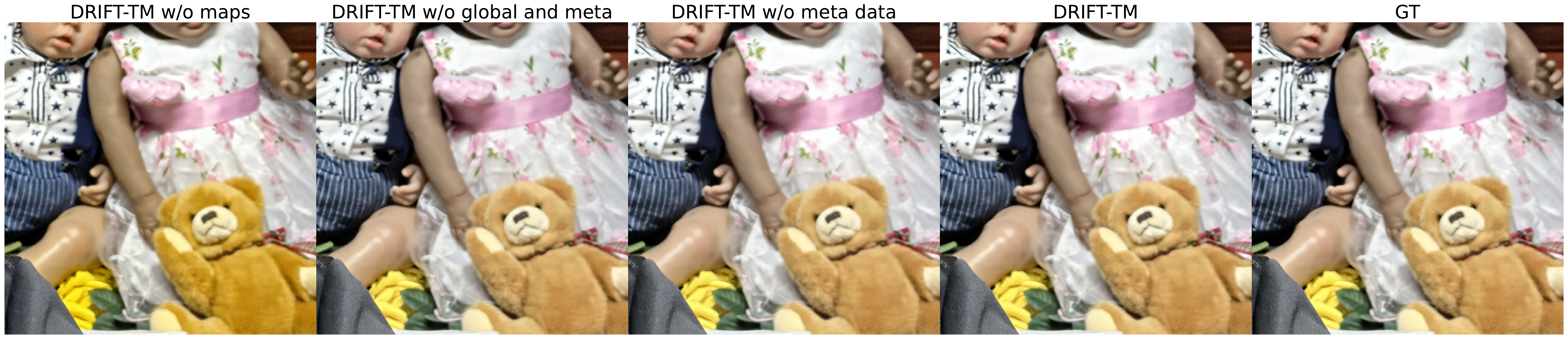}
        \label{fig:ablation_comp_image3}
        \includegraphics[width=0.9\linewidth, trim={0 1cm 0 1cm}]{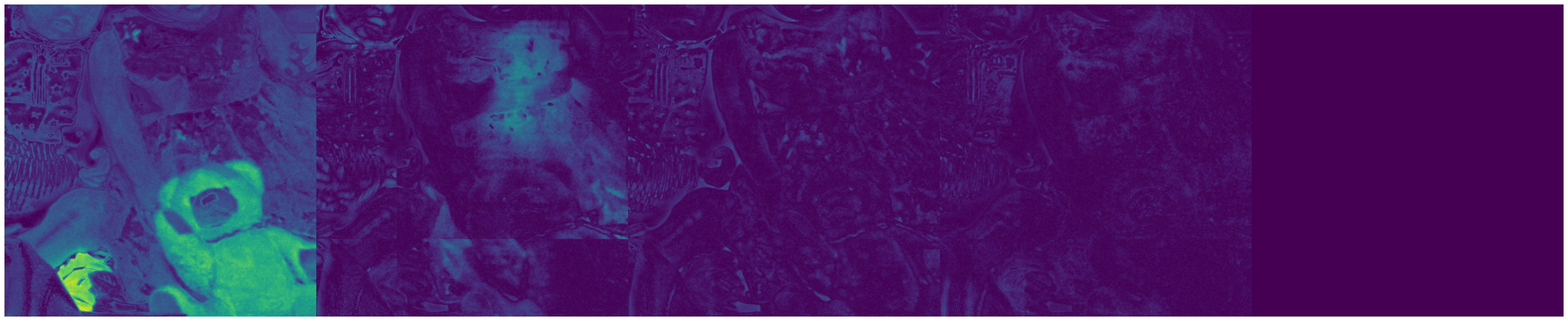}
        \label{fig:ablation_comp_image4}
\caption{Visual comparison for ablation study (top: image, bottom: difference map from GT). 
While the complete proposed method, DRIFT-TM, produces a visual match to the GT, the ablations show progressively more deviation such as tiling artifacts in "DRIFT w/o global and meta" as well as large color deviations in "DRIFT w/o maps".}
\label{fig:ablation_comp}
\end{figure}

\begin{figure}[t] 
    \centering 
        \includegraphics[width=0.8\linewidth, trim={0 1cm 0 1cm}]{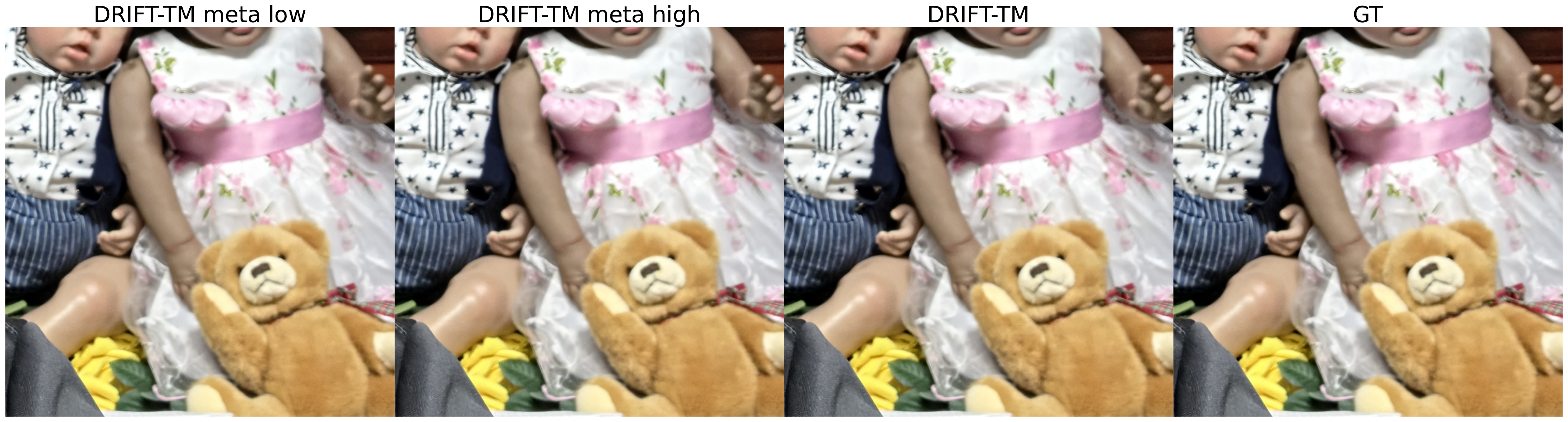}
        \label{fig:meta_comp_image3}
        \includegraphics[width=0.8\linewidth, trim={0 1cm 0 1cm}]{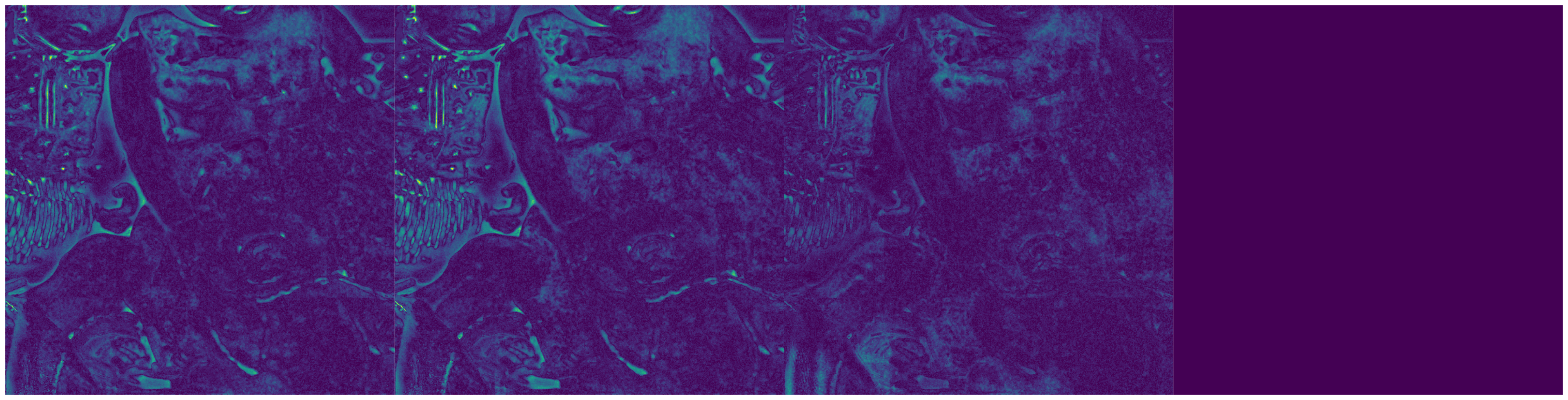}
        \label{fig:meta_comp_image4}
\caption{Visual comparison showing incorrect metadata at inference: Top (image), bottom (GT difference map). "Meta low/high" denote DRIFT-TM inference with altered ISO/exposure. 
Incorrect metadata is non-catastrophic, but accurate metadata improves tone-mapping precision.
}
\label{fig:meta_effect}
\end{figure}

We conducted an ablation study to evaluate the impact of individual components in our method, as detailed in \cref{sec:UTM}. We highlight improvements from (a) residual learning (b) using a global image information, and (c) encoding image metadata. 
Results are presented in \cref{tab:reference_results} and \cref{fig:ablation_comp} with ablations labeled as w/o maps, w/o global and meta, and w/o meta data.
Removing each feature individually, we demonstrating that all three components enhance the final image quality of DRIFT-TM.
In addition, in \cref{fig:meta_effect}, we show the effect of incorrect meta-data during inference.

\begin{figure}[t]
\centering
\begin{minipage}{0.49\linewidth}
\centering
\includegraphics[width=\linewidth, trim={0 1.2cm 0 1.2cm}]{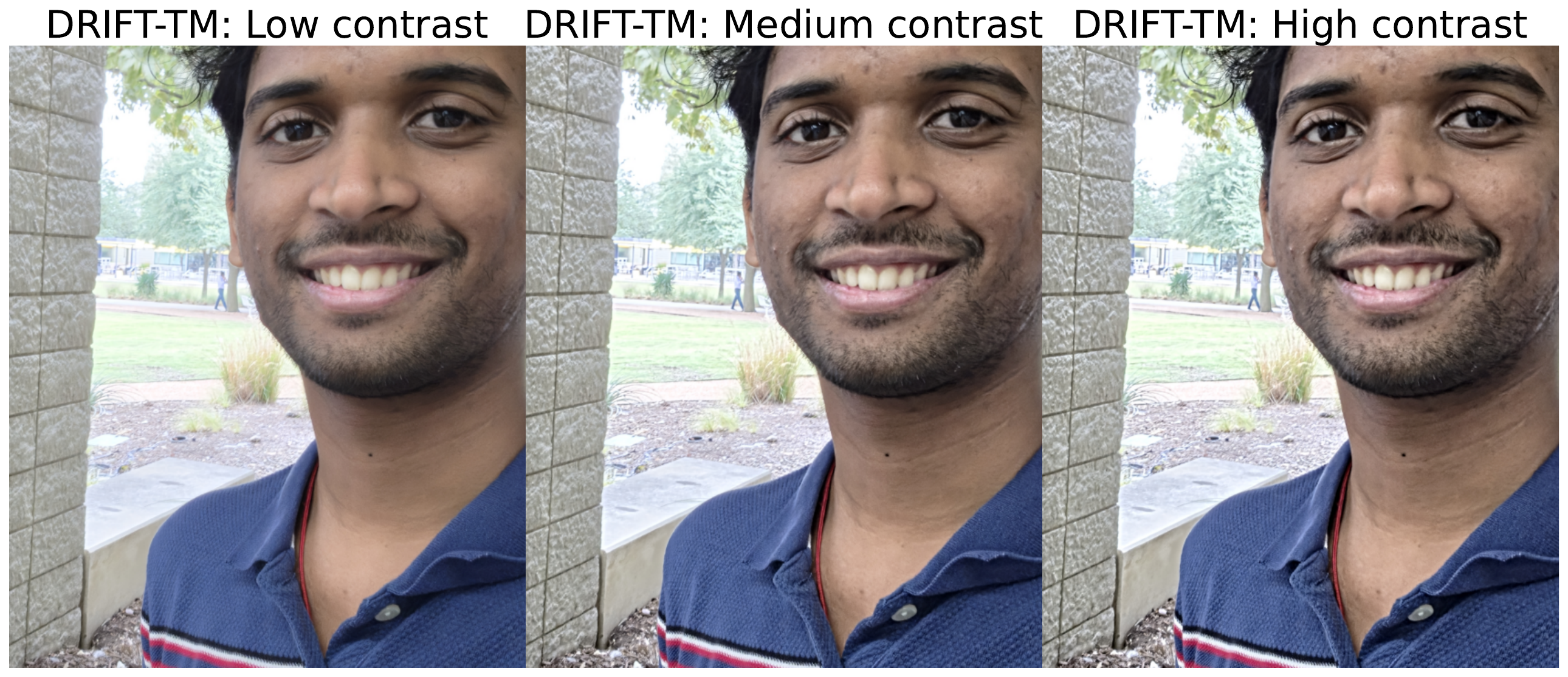}
\label{fig:tuning_image1}
\end{minipage}%
\begin{minipage}{0.49\linewidth}
\centering
\includegraphics[width=\linewidth, trim={0 1.2cm 0 1.2cm}]{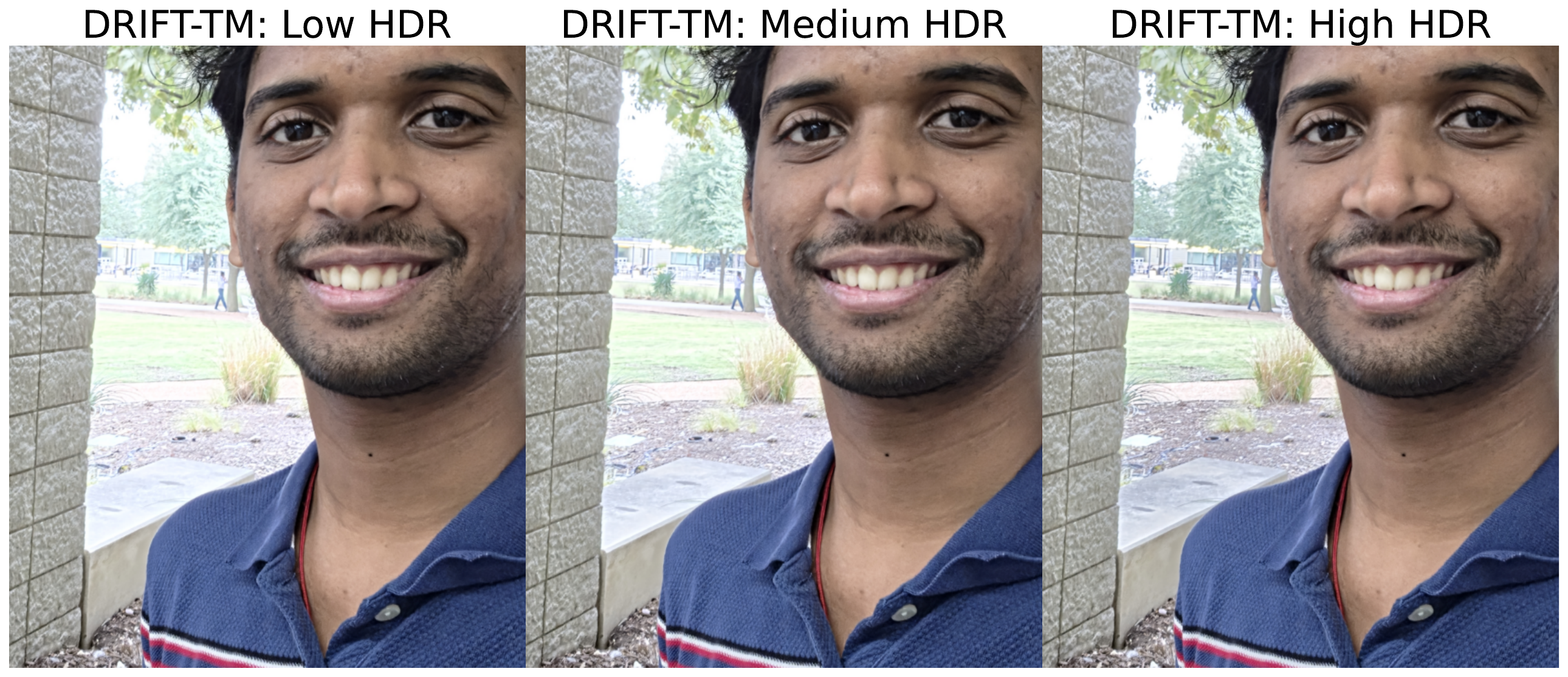}
\label{fig:tuning_image3}
\end{minipage}
\vspace{-15pt}
\caption{Illustration of our method's tunability: left shows contrast adjustment; right shows HDR effect adjustment.}
\label{fig:tuning_comp}
\end{figure}

\subsubsection{Tunability}
\label{sec:results_TM_tune}

As outlined in \cref{sec:UTM}, our method enables fine-tuning of the final output using the same trained model. This is demonstrated by adjusting tuning inputs to control contrast and HDR effects, with an example shown in \cref{fig:tuning_comp}. Further details on tunable tone enhancement are provided in \cref{sec:tunableENh}.

\subsection{Mobile Device Inference}
\label{sec:results_npu}
The NAFNet architecture used by DRIFT-MFP processes an 11-frame 12-MP resolution burst in 3.2 seconds on the Neural Processing Unit (NPU) of a Snapdragon Qualcomm 8 Elite chipset. The EV$-$ anlaysis happens in parallel on the CPU. DRIFT-Tonemap runs in 0.5 s. Since the tone-map-lite and tunable enhancement computation happens on a CPU, it runs in parallel with the tone-map network (which runs on NPU), with only the last tile compute observable. 

\section{Conclusion}
We present a novel end-to-end pipeline of deep learning-based blocks for image restoration, super-resolution, and tone-mapping that runs in less than 4s on a mobile device. Our efficient DRIFT-MFP restoration method identified and solved key shortcomings in the use of perceptual losses and was preferred in a user study. Our DRIFT-TM tone-map algorithm mimics a computationally complex algorithm while retaining key benefits of non-DL methods such as tunability and tile consistency.

%% file: main.bib
@String(CVPR= {IEEE Conf. Comput. Vis. Pattern Recog.})

@String(ECCV= {Eur. Conf. Comput. Vis.})

@String(TOG= {ACM Trans. Graph.})

@String(ICIP = {IEEE Int. Conf. Image Process.})

@String(CVPRW= {IEEE Conf. Comput. Vis. Pattern Recog. Worksh.})

@String(CVPR  = {CVPR})

@String(ECCV  = {ECCV})

@String(TOG   = {ACM TOG})

@String(ICIP  = {ICIP})

@String(CVPRW= {CVPRW})

@inproceedings{wang2018esrgan,
  title={Esrgan: Enhanced super-resolution generative adversarial networks},
  author={Wang, Xintao and Yu, Ke and Wu, Shixiang and Gu, Jinjin and Liu, Yihao and Dong, Chao and Qiao, Yu and Change Loy, Chen},
  booktitle={Proceedings of the European conference on computer vision (ECCV) workshops},
  pages={0--0},
  year={2018}
}

@article{salimans,
  title={Improved techniques for training gans},
  author={Salimans, Tim and Goodfellow, Ian and Zaremba, Wojciech and Cheung, Vicki and Radford, Alec and Chen, Xi},
  journal={Advances in neural information processing systems},
  volume={29},
  year={2016}
}

@INPROCEEDINGS{pix2pixhd,
  author={Wang, Ting-Chun and Liu, Ming-Yu and Zhu, Jun-Yan and Tao, Andrew and Kautz, Jan and Catanzaro, Bryan},
  booktitle={2018 IEEE/CVF Conference on Computer Vision and Pattern Recognition}, 
  title={High-Resolution Image Synthesis and Semantic Manipulation with Conditional GANs}, 
  year={2018},
  volume={},
  number={},
  pages={8798-8807},
  doi={10.1109/CVPR.2018.00917}}

@article{xu_multi-exposure_2022,
	title = {Multi-{Exposure} {Image} {Fusion} {Techniques}: {A} {Comprehensive} {Review}},
	volume = {14},
	copyright = {http://creativecommons.org/licenses/by/3.0/},
	issn = {2072-4292},
	shorttitle = {Multi-{Exposure} {Image} {Fusion} {Techniques}},
	url = {https://www.mdpi.com/2072-4292/14/3/771},
	doi = {10.3390/rs14030771},
	language = {en},
	number = {3},
	urldate = {2024-12-31},
	journal = {Remote Sensing},
	author = {Xu, Fang and Liu, Jinghong and Song, Yueming and Sun, Hui and Wang, Xuan},
	month = jan,
	year = {2022},
	note = {Number: 3
Publisher: Multidisciplinary Digital Publishing Institute},
	pages = {771},
}

@article{ponomarenko_image_2015,
	title = {Image database {TID2013}: {Peculiarities}, results and perspectives},
	volume = {30},
	issn = {09235965},
	shorttitle = {Image database {TID2013}},
	url = {https://linkinghub.elsevier.com/retrieve/pii/S0923596514001490},
	doi = {10.1016/j.image.2014.10.009},
	language = {en},
	urldate = {2024-12-24},
	journal = {Signal Processing: Image Communication},
	author = {Ponomarenko, Nikolay and Jin, Lina and Ieremeiev, Oleg and Lukin, Vladimir and Egiazarian, Karen and Astola, Jaakko and Vozel, Benoit and Chehdi, Kacem and Carli, Marco and Battisti, Federica and Jay Kuo, C.-C.},
	month = jan,
	year = {2015},
	pages = {57--77},
}

@inproceedings{dudhane2022burst,
  title={Burst image restoration and enhancement},
  author={Dudhane, Akshay and Zamir, Syed Waqas and Khan, Salman and Khan, Fahad Shahbaz and Yang, Ming-Hsuan},
  booktitle={Proceedings of the ieee/cvf Conference on Computer Vision and Pattern Recognition},
  pages={5759--5768},
  year={2022}
}

@inproceedings{dudhane2023burstormer,
  title={Burstormer: Burst image restoration and enhancement transformer},
  author={Dudhane, Akshay and Zamir, Syed Waqas and Khan, Salman and Khan, Fahad Shahbaz and Yang, Ming-Hsuan},
  booktitle={2023 IEEE/CVF Conference on Computer Vision and Pattern Recognition (CVPR)},
  pages={5703--5712},
  year={2023},
  organization={IEEE}
}

@article{wronski2019handheld,
  title={Handheld multi-frame super-resolution},
  author={Wronski, Bartlomiej and Garcia-Dorado, Ignacio and Ernst, Manfred and Kelly, Damien and Krainin, Michael and Liang, Chia-Kai and Levoy, Marc and Milanfar, Peyman},
  journal={ACM Transactions on Graphics (ToG)},
  volume={38},
  number={4},
  pages={1--18},
  year={2019},
  publisher={ACM New York, NY, USA}
}

@inproceedings{tico2008multi,
  title={Multi-frame image denoising and stabilization},
  author={Tico, Marius},
  booktitle={2008 16th European Signal Processing Conference},
  pages={1--4},
  year={2008},
  organization={IEEE}
}

@inproceedings{khan2025mfsr,
  title={MFSR-GAN: Multi-Frame Super-Resolution with Handheld Motion Modeling},
  author={Khan, Fadeel Sher and Ebenezer, Joshua and Sheikh, Hamid and Lee, Seok-Jun},
  booktitle={Proceedings of the Computer Vision and Pattern Recognition Conference},
  pages={800--809},
  year={2025}
}

@inproceedings{johnson2016perceptual,
  title={Perceptual losses for real-time style transfer and super-resolution},
  author={Johnson, Justin and Alahi, Alexandre and Fei-Fei, Li},
  booktitle={European conference on computer vision},
  pages={694--711},
  year={2016},
  organization={Springer}
}

@article{goodfellow2014generative,
  title={Generative adversarial nets},
  author={Goodfellow, Ian J and Pouget-Abadie, Jean and Mirza, Mehdi and Xu, Bing and Warde-Farley, David and Ozair, Sherjil and Courville, Aaron and Bengio, Yoshua},
  journal={Advances in neural information processing systems},
  volume={27},
  year={2014}
}

@inproceedings{isola2017image,
  title={Image-to-image translation with conditional adversarial networks},
  author={Isola, Phillip and Zhu, Jun-Yan and Zhou, Tinghui and Efros, Alexei A},
  booktitle={Proceedings of the IEEE conference on computer vision and pattern recognition},
  pages={1125--1134},
  year={2017}
}

@article{ANAYA2018144,
title = {RENOIR – A dataset for real low-light image noise reduction},
journal = {Journal of Visual Communication and Image Representation},
volume = {51},
pages = {144-154},
year = {2018},
issn = {1047-3203},
doi = {https://doi.org/10.1016/j.jvcir.2018.01.012},
url = {https://www.sciencedirect.com/science/article/pii/S1047320318300208},
author = {Josue Anaya and Adrian Barbu}
}

@misc{snpe,
  author={{Qualcomm}},
  title = {{Qualcomm AI Runtime SDK}},
  howpublished = {\url{https://docs.qualcomm.com/bundle/publicresource/topics/80-63442-10/SNPE_general_revision_history.html,}},
  note = {Accessed: 2025-11-13},
  year={2025},
}

@INPROCEEDINGS{ntire2025night,
  author={Ershov, Egor and others},
  booktitle={2025 IEEE/CVF Conference on Computer Vision and Pattern Recognition Workshops (CVPRW)}, 
  title={NTIRE 2025 Challenge on Night Photography Rendering}, 
  year={2025},
  volume={},
  number={},
  pages={1505-1515},
  doi={10.1109/CVPRW67362.2025.00140}}

@incollection{avidan_simple_2022,
	address = {Cham},
	title = {Simple {Baselines} for {Image} {Restoration}},
	volume = {13667},
	isbn = {978-3-031-20070-0 978-3-031-20071-7},
	url = {https://link.springer.com/10.1007/978-3-031-20071-7_2},
	language = {en},
	urldate = {2024-12-24},
	booktitle = {Computer {Vision} – {ECCV} 2022},
	publisher = {Springer Nature Switzerland},
	author = {Chen, Liangyu and Chu, Xiaojie and Zhang, Xiangyu and Sun, Jian},
	editor = {Avidan, Shai and Brostow, Gabriel and Cissé, Moustapha and Farinella, Giovanni Maria and Hassner, Tal},
	year = {2022},
	doi = {10.1007/978-3-031-20071-7_2},
	note = {Series Title: Lecture Notes in Computer Science},
	pages = {17--33},
}

@inproceedings{zhang2018unreasonable,
  title={The unreasonable effectiveness of deep features as a perceptual metric},
  author={Zhang, Richard and Isola, Phillip and Efros, Alexei A and Shechtman, Eli and Wang, Oliver},
  booktitle={Proceedings of the IEEE conference on computer vision and pattern recognition},
  pages={586--595},
  year={2018}
}

@article{simonyan2014very,
  title={Very deep convolutional networks for large-scale image recognition},
  author={Simonyan, Karen and Zisserman, Andrew},
  journal={arXiv preprint arXiv:1409.1556},
  year={2014}
}

@INPROCEEDINGS{imagenet,
  author={Deng, Jia and Dong, Wei and Socher, Richard and Li, Li-Jia and Kai Li and Li Fei-Fei},
  booktitle={2009 IEEE Conference on Computer Vision and Pattern Recognition}, 
  title={ImageNet: A large-scale hierarchical image database}, 
  year={2009},
  volume={},
  number={},
  pages={248-255},
  doi={10.1109/CVPR.2009.5206848}}

@InProceedings{Bhat_2021_CVPR,
    author    = {Bhat, Goutam and Danelljan, Martin and Van Gool, Luc and Timofte, Radu},
    title     = {Deep Burst Super-Resolution},
    booktitle = {Proceedings of the IEEE/CVF Conference on Computer Vision and Pattern Recognition (CVPR)},
    month     = {June},
    year      = {2021},
    pages     = {9209-9218}
}

@InProceedings{Zamir_2022_CVPR,
    author    = {Zamir, Syed Waqas and Arora, Aditya and Khan, Salman and Hayat, Munawar and Khan, Fahad Shahbaz and Yang, Ming-Hsuan},
    title     = {Restormer: Efficient Transformer for High-Resolution Image Restoration},
    booktitle = {Proceedings of the IEEE/CVF Conference on Computer Vision and Pattern Recognition (CVPR)},
    month     = {June},
    year      = {2022},
    pages     = {5728-5739}
}

@InProceedings{Zamir_2021_CVPR,
    author    = {Zamir, Syed Waqas and Arora, Aditya and Khan, Salman and Hayat, Munawar and Khan, Fahad Shahbaz and Yang, Ming-Hsuan and Shao, Ling},
    title     = {Multi-Stage Progressive Image Restoration},
    booktitle = {Proceedings of the IEEE/CVF Conference on Computer Vision and Pattern Recognition (CVPR)},
    month     = {June},
    year      = {2021},
    pages     = {14821-14831}
}

@article{llflut,
  title={High-resolution Photo Enhancement in Real-time: A Laplacian Pyramid Network},
  author={Zhang, Feng and Deng, Haoyou and Li, Zhiqiang and Li, Lida and Xu, Bin and Lu, Qingbo and Cao, Zisheng and Wei, Minchen and Gao, Changxin and Sang, Nong and others},
  journal={IEEE Transactions on Pattern Analysis and Machine Intelligence},
  year={2025},
  publisher={IEEE}
}

@inproceedings{tmogan,
  title={A generative adversarial network based tone mapping operator for 4k hdr images},
  author={Zhang, Junbin and Wang, Yixiao and Tohidypour, Hamidreza and Pourazad, Mahsa T and Nasiopoulos, Panos},
  booktitle={2023 international conference on computing, networking and communications (ICNC)},
  pages={473--477},
  year={2023},
  organization={IEEE}
}

@ARTICLE{yeganeh2013Objective,
  author={Yeganeh, Hojatollah and Wang, Zhou},
  journal={IEEE Transactions on Image Processing}, 
  title={Objective Quality Assessment of Tone-Mapped Images}, 
  year={2013},
  volume={22},
  number={2},
  pages={657-667},
  doi={10.1109/TIP.2012.2221725}
}

@inproceedings{wang2022learning,
  title={Learning a self-supervised tone mapping operator via feature contrast masking loss},
  author={Wang, Chao and Chen, Bin and Seidel, Hans-Peter and Myszkowski, Karol and Serrano, Ana},
  booktitle={Computer Graphics Forum},
  volume={41},
  number={2},
  pages={71--84},
  year={2022},
  organization={Wiley Online Library}
}

@ARTICLE{guo2021deep,
  author={Guo, Cheng and Jiang, Xiuhua},
  journal={IEEE Access}, 
  title={Deep Tone-Mapping Operator Using Image Quality Assessment Inspired Semi-Supervised Learning}, 
  year={2021},
  volume={9},
  number={},
  pages={73873-73889},
  doi={10.1109/ACCESS.2021.3080331}}

@inproceedings{conde2025ntire,
  title={NTIRE 2025 challenge on raw image restoration and super-resolution},
  author={Conde, Marcos and others},
  booktitle={Proceedings of the Computer Vision and Pattern Recognition Conference},
  pages={1148--1171},
  year={2025}
}

@inproceedings{lee2025ntire,
  title={NTIRE 2025 challenge on efficient burst hdr and restoration: Datasets, methods, and results},
  author={Lee, Sangmin and others},
  booktitle={Proceedings of the Computer Vision and Pattern Recognition Conference},
  pages={1002--1017},
  year={2025}
}

@article{mantiuk2008display,
author = {Mantiuk, Rafa\l{} and Daly, Scott and Kerofsky, Louis},
title = {Display adaptive tone mapping},
year = {2008},
issue_date = {August 2008},
publisher = {Association for Computing Machinery},
address = {New York, NY, USA},
volume = {27},
number = {3},
issn = {0730-0301},
url = {https://doi.org/10.1145/1360612.1360667},
doi = {10.1145/1360612.1360667},
journal = {ACM Trans. Graph.},
month = aug,
pages = {1–10},
numpages = {10}
}

@incollection{reinhard2021high,
  title={High dynamic range imaging},
  author={Reinhard, Erik},
  booktitle={Computer Vision: A Reference Guide},
  pages={558--563},
  year={2021},
  publisher={Springer}
}

@article{gastal2011domain,
author = {Gastal, Eduardo S. L. and Oliveira, Manuel M.},
title = {Domain transform for edge-aware image and video processing},
year = {2011},
issue_date = {July 2011},
publisher = {Association for Computing Machinery},
address = {New York, NY, USA},
volume = {30},
number = {4},
issn = {0730-0301},
url = {https://doi.org/10.1145/2010324.1964964},
doi = {10.1145/2010324.1964964},
journal = {ACM Trans. Graph.},
month = jul,
articleno = {69},
numpages = {12}
}

@article{petit2013assessment,
  title={Assessment of video tone-mapping: Are cameras’ S-shaped tone-curves good enough?},
  author={Petit, Josselin and Mantiuk, Rafa{\l} K},
  journal={Journal of Visual Communication and Image Representation},
  volume={24},
  number={7},
  pages={1020--1030},
  year={2013},
  publisher={Elsevier}
}

@inproceedings{mertens2007exposure,
  title={Exposure fusion},
  author={Mertens, Tom and Kautz, Jan and Van Reeth, Frank},
  booktitle={15th Pacific Conference on Computer Graphics and Applications (PG'07)},
  pages={382--390},
  year={2007},
  organization={IEEE}
}

@article{fu2022edge,
  title={Edge-aware deep image deblurring},
  author={Fu, Zhichao and Zheng, Yingbin and Ma, Tianlong and Ye, Hao and Yang, Jing and He, Liang},
  journal={Neurocomputing},
  volume={502},
  pages={37--47},
  year={2022},
  publisher={Elsevier}
}

@article{Krawczyk_2023,
title={Artifact generation when using perceptual loss for image deblurring},
url={http://dx.doi.org/10.36227/techrxiv.23791962.v1},
DOI={10.36227/techrxiv.23791962.v1},
publisher={Institute of Electrical and Electronics Engineers (IEEE)},
author={Krawczyk, Patrick and Gaertner, Marvin and Jansche, Andreas and Bernthaler, Timo and Schneider, Gerhard},
journal={TechRxiv},
year={2023},
month=jul }

@misc{goodfellow2016deep,
  title={Deep learning},
  author={Goodfellow, Ian},
  year={2016},
  publisher={MIT press}
}

@article{pihlgren2023systematic,
  title={A systematic performance analysis of deep perceptual loss networks: Breaking transfer learning conventions},
  author={Pihlgren, Gustav Grund and Nikolaidou, Konstantina and Chhipa, Prakash Chandra and Abid, Nosheen and Saini, Rajkumar and Sandin, Fredrik and Liwicki, Marcus},
  journal={arXiv preprint arXiv:2302.04032},
  year={2023}
}

@inproceedings{wei2023towards,
  title={Towards real-world burst image super-resolution: Benchmark and method},
  author={Wei, Pengxu and Sun, Yujing and Guo, Xingbei and Liu, Chang and Li, Guanbin and Chen, Jie and Ji, Xiangyang and Lin, Liang},
  booktitle={Proceedings of the IEEE/CVF International Conference on Computer Vision},
  pages={13233--13242},
  year={2023}
}

@article{cerda2018tone,
  title={Which tone-mapping operator is the best? A comparative study of perceptual quality},
  author={Cerda-Company, Xim and Parraga, C Alejandro and Otazu, Xavier},
  journal={Journal of the Optical Society of America A},
  volume={35},
  number={4},
  pages={626--638},
  year={2018},
  publisher={Optical Society of America}
}

@article{delbracio2021mobile,
  title={Mobile computational photography: A tour},
  author={Delbracio, Mauricio and Kelly, Damien and Brown, Michael S and Milanfar, Peyman},
  journal={Annual review of vision science},
  volume={7},
  number={1},
  pages={571--604},
  year={2021},
  publisher={Annual Reviews}
}

@inproceedings{kinoshita2019convolutional,
  title={Convolutional neural networks considering local and global features for image enhancement},
  author={Kinoshita, Yuma and Kiya, Hitoshi},
  booktitle={2019 IEEE International Conference on Image Processing (ICIP)},
  pages={2110--2114},
  year={2019},
  organization={IEEE}
}

@article{rana2019deep,
  title={Deep tone mapping operator for high dynamic range images},
  author={Rana, Aakanksha and Singh, Praveer and Valenzise, Giuseppe and Dufaux, Frederic and Komodakis, Nikos and Smolic, Aljosa},
  journal={IEEE Transactions on Image Processing},
  volume={29},
  pages={1285--1298},
  year={2019},
  publisher={IEEE}
}

@article{ma2015high,
  title={High dynamic range image compression by optimizing tone mapped image quality index},
  author={Ma, Kede and Yeganeh, Hojatollah and Zeng, Kai and Wang, Zhou},
  journal={IEEE Transactions on Image Processing},
  volume={24},
  number={10},
  pages={3086--3097},
  year={2015},
  publisher={IEEE}
}

@article{yang2024learning,
  title={Learning differential pyramid representation for tone mapping},
  author={Yang, Qirui and Li, Yinbo and Liu, Yihao and Jiang, Peng-Tao and Zhang, Fangpu and Cheng, Qihua and Yue, Huanjing and Yang, Jingyu},
  journal={arXiv preprint arXiv:2412.01463},
  year={2024}
}

@article{zhang2023lookup,
  title={Lookup table meets local laplacian filter: pyramid reconstruction network for tone mapping},
  author={Zhang, Feng and Tian, Ming and Li, Zhiqiang and Xu, Bin and Lu, Qingbo and Gao, Changxin and Sang, Nong},
  journal={Advances in Neural Information Processing Systems},
  volume={36},
  pages={57558--57569},
  year={2023}
}

@article{fischler1981random,
  title={Random sample consensus: a paradigm for model fitting with applications to image analysis and automated cartography},
  author={Fischler, Martin A and Bolles, Robert C},
  journal={Communications of the ACM},
  volume={24},
  number={6},
  pages={381--395},
  year={1981}
}

@article{burt1983multiresolution,
  title={A multiresolution spline with application to image mosaics},
  author={Burt, Peter J and Adelson, Edward H},
  journal={ACM Transactions on Graphics (ToG)},
  volume={2},
  number={4},
  pages={217--236},
  year={1983},
  publisher={ACM New York, NY, USA}
}

@inproceedings{mertens2009exposure,
  title={Exposure fusion: A simple and practical alternative to high dynamic range photography},
  author={Mertens, Tom and Kautz, Jan and Van Reeth, Frank},
  booktitle={Computer graphics forum},
  volume={28},
  number={1},
  pages={161--171},
  year={2009},
  organization={Wiley Online Library}
}

@inproceedings{madhusudana2024mobile,
  title={Mobile Aware Denoiser Network (MADNet) for Quad Bayer Images},
  author={Madhusudana, Pavan C and Li, Jing and Nadir, Zeeshan and Sheikh, Hamid R and Lee, Seok-Jun},
  booktitle={Proceedings of the IEEE/CVF Conference on Computer Vision and Pattern Recognition},
  pages={44--52},
  year={2024}
}

@inproceedings{li2019hybrid,
  title={Hybrid synthesis for exposure fusion from hand-held camera inputs},
  author={Li, Ru and Liu, Shuaicheng and Liu, Guanghui and Zeng, Bing},
  booktitle={2019 IEEE International Conference on Image Processing (ICIP)},
  pages={4639--4643},
  year={2019},
  organization={IEEE}
}

@article{yang2025high,
  title={High dynamic range image tone mapping based on variational image decomposition and color correction},
  author={Yang, Xuejie and Zheng, Huamiao and Su, Yonggang},
  journal={Optics \& Laser Technology},
  volume={181},
  pages={111873},
  year={2025},
  publisher={Elsevier}
}
